\documentclass[10pt]{emulateapj}
\usepackage{amsmath}
\usepackage{graphicx}
\usepackage{amssymb}
\usepackage{times}
\usepackage{color}

\def\gtorder{\mathrel{\raise.3ex\hbox{$>$}\mkern-14mu
             \lower0.6ex\hbox{$\sim$}}}
\def\ltorder{\mathrel{\raise.3ex\hbox{$<$}\mkern-14mu
             \lower0.6ex\hbox{$\sim$}}}

\def\na{New Ast.\,}

\slugcomment{Draft of \today}
\shorttitle{Late time light curve of GW\,170817}
\shortauthors{Waxman et al.}

\begin{document}

\title{Late time kilonova light curves and implications to GW\,170817}
\author{Eli Waxman\altaffilmark{1}, Eran O. Ofek\altaffilmark{1}, Doron Kushnir\altaffilmark{1}}
\altaffiltext{1}{Dept. of Particle Phys. \& Astrophys., Weizmann Institute of Science, Rehovot 76100, Israel}
\begin{abstract}
We discuss the late time (tens of days) emission from the radioactive ejecta of mergers involving neutron stars, when the ionization energy loss time of beta-decay electrons and positrons exceeds the expansion time. We show that if the e$^\pm$ are confined to the plasma (by magnetic fields), then the time dependence of the plasma heating rate, $\dot{\varepsilon}_d$, and hence of the bolometric luminosity $L=\dot{\varepsilon}_d$, are given by $d\log L/d\log t\simeq-2.8$, nearly independent of the composition and of the instantaneous radioactive energy release rate, $\dot{\varepsilon}$.
This universality of the late time behavior is due to the weak dependence of the ionization loss rate on composition and on e$^\pm$ energy.
The late time IR and optical measurements of GW\,170817 are consistent with this expected behavior provided that the ionization loss time exceeds the expansion time at $t>t_\varepsilon\approx 7$~d, as predicted based on the early (few day) electromagnetic emission.
\end{abstract}

\maketitle

\section{Introduction}
\label{sec:Intro}

The merger of a neutron star with its binary neutron star or black hole companion has been suggested \citep{LattimerSchramm74} to produce high density neutron rich ejecta, in which heavy elements beyond Iron are produced by the r-process. Heating of the expanding ejecta by radioactive decay of unstable isotopes was in turn predicted \citep{LiPac98} to produce a strong optical-UV emission, commonly referred to as a "kilonova" \citep[see e.g.][for a recent review]{2016FernandezMetzgerRev}. Remarkably, the observed UV-IR emission following the neutron star merger event GW\,170817 \citep{2017Sci...358.1556C,2017ApJ...848L..16S,2017ApJ...848L..24V,Arcavi2017-Discovery,Lipunov2018,Tanvir17}, which was detected through its gravitational wave emission \citep{Abbott17PhRvL}, is broadly consistent with these predictions \citep[e.g.][]{Kasen17Nat,2017DroutKN-LC,Rosswog17,CowperthwaiteBerger173KN,Tanaka17,PeregoRadice173KN,Kilpatrick17,2018WOKG}.
However, our understanding of the UV-IR signal and of the constraints it provides on the structure and composition of the ejecta is incomplete.

The early UV/blue emission suggests the existence of a fairly massive, few $\times 10^{-2}M_\odot$, and fast, $v\sim 0.3c$, component of the ejecta with low opacity,
$\kappa<1$\,cm$^{2}$\,g$^{-1}$, corresponding to a large initial electron fraction $Y_e$ leading to a small fraction in the ejecta of $r$-process elements beyond $A\simeq140$, which are expected to have much larger opacity \citep[e.g.][]{Kasen13}. The presence of this massive component is in tension with the results of detailed numeric merger calculations, that generally predict high opacity for the massive, fast ejecta components. This has lead to a discussion of possible mechanisms for the generation of the observed low opacity ejecta \citep[e.g.][]{2018MetzgerQuataertMagnetarKN,2018FujibayashiViscousEjecta}. Alternative models have also been proposed, in which the blue emission is produced by boosted relativistic material and/or shock cooling of an expanding mildly relativistic shell \citep[e.g.][]{2017Kasliwal,Piro17,Gottlieb18Cocoon}.

On a few days time scale, the peak emission shifted from the UV/blue to the red/IR. This observed blue to red evolution may be explained by the existence of several ejecta components,
characterized by largely differing compositions, with higher opacity components dominating at later times \citep[e.g.][]{Kasen17Nat,2017DroutKN-LC,Rosswog17,CowperthwaiteBerger173KN,Tanaka17,PeregoRadice173KN,Kilpatrick17}. Using a simple analytic model for the UV-IR emission we have shown \citep[][hereafter Paper I]{2018WOKG} that an alternative explanation is possible, in which the entire ejecta is composed of low opacity, $\kappa<1$\,cm$^{2}$\,g$^{-1}$, material \citep[see also][]{2017Natur.551...75S,Rosswog17}.
The inferred significant opacity at the 1-2~$\mu m$ band at a few days, $\kappa\approx0.1$\,cm$^{2}$\,g$^{-1}$, provides an important constraint on the composition, the implication of which is uncertain due to the uncertainty in the opacity (see discussion in Paper I).

Our model deviates from those of earlier work in two aspects, which are key to enabling an explanation of all data \citep[see][]{2018WOKG,Arcavi2018-EarlyEM} with a nearly uniform composition ejecta: allowing a wider velocity distribution than previously assumed, and including the suppression of radioactive heating due to adiabatic losses of beta-decay electrons and positrons. Hereafter, we use "beta-decay electrons" to refer to both beta-decay electrons and positrons.

On a time scale of minutes to tens of days the main radioactive energy source is expected to be beta-decays \citep[alpha-decays may provide a significant contribution on a 100~d time scale, e.g.][]{Barneskasen16}.
While neutrinos (and, at this stage, also $\gamma$-rays) escape the ejecta, the electrons produced in beta-decays heat it through ionization and plasma losses. As the ejecta expand, the electron ionization loss time increases in proportion to $\rho^{-1}\propto t^{3}$, exceeding the expansion time $t$ at $t>t_\varepsilon$ (the expansion and loss time are equal at $t=t_{\epsilon}$).
We have pointed out in Paper I that while the beta-decay electrons are likely confined (by magnetic fields\footnote{Magnetic confinement appears likely since it would be facilitated by a relatively weak magnetic field. Assuming $B\propto r^{-2}$ in the expanding ejecta, as may be appropriate for a tangled field, $B\gtrsim0.1\mu$G is expected at $t\sim10$\,d, $r\sim10^{16}$\,cm (the equipartition field at 10\,d is $\sim1$\,mG), implying an electron Larmor radius of $\sim10^{10}$\,cm$\ll r\sim10^{16}$.}) to the plasma, the fraction of their energy that is converted to heat at $t>t_\varepsilon$ is only $(t/t_\varepsilon)^{-2}$, since they lose energy to adiabatic expansion (applying pressure against the expanding plasma) faster than to ionization. This leads to a steepening of the decline of the bolometric luminosity at $t\sim t_\varepsilon$. Note that adiabatic losses become important for supernovae much later than in the kilonova case due to the larger mass and slower velocity of the ejecta of supernovae, $t_\varepsilon\propto M^{1/2}v^{-3/2}$, see Eq.~(\ref{eq:t_E}).

Analyzing the UV-IR data of GW\,170817, we have shown in Paper I that the analytic model for the emission by a uniform composition ejecta is over-constrained. All its parameters are determined by the observations at $t<6$\,d, predicting a transition to inefficient energy deposition and transparency at roughly the same time, $t=t_\varepsilon\approx 7$~d, and $L\propto t^{-3}$ at $t>t_\varepsilon$. This is consistent with the observed steepening of the light curve at $\approx7$\,d, and with the strong deviation from thermal spectra at $t\gtorder5$\,d.

We return here to the issue of the late time kilonova light curve for two reasons. First, new IR and optical observations extend the range of time over which data are available to $>100$~d \citep{2018Kasliwal_IR,Lamb2019}, compared to $\sim15$~d at the time Paper I was written. This enables further tests of the models. Second, in Paper I we have neglected plasma heating by low energy electrons, that may accumulate in the ejecta if they are confined by magnetic fields. \citet{2018KB_dEdX} have pointed out that such accumulation leads to a shallower decline of the bolometric luminosity at $t>t_\varepsilon$, in particular if the ionization energy loss rate increases with decreasing electron energy. We improve our analytic model to include these effects. Our results differ from those of \citet{2018KB_dEdX} (who find $L=\dot{\varepsilon}_d\propto\dot{\varepsilon} t^{-1}$),
for reasons explained in \S~\ref{sec:theory} (see end of \S~\ref{sec:analytic_lc}).

This paper is organized as follows. The processes affecting the late time, $t\gtrsim t_\varepsilon$, radioactive heating of the ejecta are discussed in \S~\ref{sec:late_rad_heat}. A simple analytic derivation of the asymptotic, $t\gg t_\varepsilon$, light curve behavior for the case where electrons are confined to the plasma is given in \S~\ref{sec:asym}. The range of validity of the asymptotic behavior is discussed based on a comparison to complete solutions of the equations (given in the appendix). The main results of the analysis of \S~\ref{sec:late_rad_heat} and \S~\ref{sec:asym} are given in \S~\ref{sec:analytic_lc}. Simple analytic formulae describing the late-time kilonova light curves, updating Eq.~(12) of Paper I to include the effects of the accumulation of low-energy electrons and of the energy dependence of the ionization losses, are given. The new, late time, observations of GW\,170817 are analyzed in \S~\ref{sec:GW170817}. Our conclusions are discussed in \S~\ref{sec:discussion}.

\section{Late time kilonova light curves}
\label{sec:theory}

\subsection{Late radioactive heating}
\label{sec:late_rad_heat}

We consider a fluid element within the expanding ejecta, with velocity $v$ and density $\rho\propto t^{-3}$. As noted in the introduction, on a time scale of minutes to tens of days the main radioactive energy source is expected to be beta-decays, while alpha-decays may provide a significant contribution on a 100~d time scale. We consider first heating by beta-decay, and return to alpha-decay at the end of this sub-section. Since neutrinos and, at the late times in which we are interested here, also gamma-rays escape the plasma, we define $\dot{\varepsilon}$ as the rate at which radioactive energy is released in the form of  electron kinetic energy.

The electrons and positrons lose energy to plasma heating mainly by ionization \citep[e.g.][]{Longair92Book},
\begin{eqnarray}\label{eq:dEdX}
  \frac{dE}{dX}&=&\frac{4\pi e^4}{m_e m_p v_e^2}\frac{Z}{A} \Big[\ln\left(\frac{\gamma_e^2 m_e v_e^2}{\bar{I}}\right) -\frac{1}{2}\ln(1+\gamma_e)\nonumber \\
  &-&\left(\frac{2\gamma_e+\gamma_e^2-1}{2\gamma_e^2}\right)\ln2+\frac{1}{2\gamma_e^2}
  +\frac{1}{16}\left(1-\frac{1}{\gamma_e}\right)^2\Big],
\end{eqnarray}
where $dE/dX$ is the energy loss per unit column density (grammage) traversed by the electron/positron, $dX=\rho dx$, $Z$ and $A$ are the atomic and mass numbers of the plasma nuclei, and $\bar{I}$ is the effective average ionization energy of the atoms, which is empirically determined and approximately given by $\bar{I}=10Z$\,eV for heavy nuclei.

Electrons lose energy also by scattering free plasma electrons ("plasma losses"). The plasma loss rate is given by an equation similar to Eq.~(\ref{eq:dEdX}), the main differences being replacing $Z$ with the number of free electrons per atom and $\bar{I}$ with $\hbar\omega_p$, where $\omega_p$ is the plasma frequency.
At the times under discussion, the ejecta temperature is a fraction of 1\,eV and its number density is $n\sim10^{6}(t/10{\rm d})^{-3}{\rm cm^{-3}}$ of $A\sim100$ atoms.
Under these conditions, the losses are dominated by ionization for electron kinetic energy larger than $\bar{I}\sim1$~keV (for $\hbar\omega_p\sim10^{-7}$~eV and ionization of a few, plasma losses at $E\sim1$~MeV amount to $\lesssim20\%$ of the ionization losses for high $Z$ nuclei). Finally, at highly relativistic energy Bremsstrahlung losses dominate over ionization \citep[e.g.][]{Longair92Book},
\begin{eqnarray}\label{eq:dEdX_Bremm}
  \frac{dE^B}{dX}&=&\frac{4e^4}{m_e m_p c v_e}\frac{Z^2e^2}{A\hbar c}\frac{E}{m_ec^2} \Big[\ln\left(\frac{183}{Z^{1/3}}\right)+\frac{1}{8}\Big].
\end{eqnarray}
Note that since gamma-rays escape the plasma, only a fraction of the Bremsstrahlung energy is deposited in the plasma.

\begin{figure}
\centerline{\includegraphics[width=8cm]{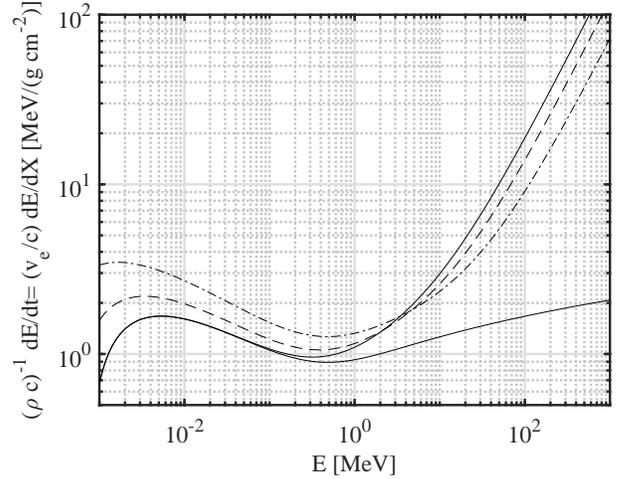}}
\caption{Energy loss rates (by ionization and Bremsstrahlung) as a function of electron kinetic energy, $E=(\gamma_e-1)m_ec^2$. The solid, dashed, and dash-dotted lines show the loss rates, given by Eqs.~(\ref{eq:dEdX}) and~(\ref{eq:dEdX_Bremm}) (with $\bar{I}=10Z$~eV) for propagation trough a plasma of $^{207}_{82}$Pb, $^{127}_{53}$I and $^{56}_{26}$Fe respectively. The lower solid curve shows the ionization loss rate for Pb. The ionization loss rate is nearly independent of $E$ and of the plasma composition.\label{fig:dEdX}}
\end{figure}

Figure~\ref{fig:dEdX} shows the energy loss rate of electrons as a function of their kinetic energy, $E=(\gamma_e-1)m_ec^2$. The figure demonstrates that the ionization loss rate is nearly independent of $E$ and of the plasma composition. Following Paper I, we define an effective ionization "opacity",
\begin{equation}\label{eq:kappa_e}
  \kappa_e \equiv E(dE/dX)^{-1}\big|_{E=1{\rm MeV}}\approx 1 {\rm cm^2/g},
\end{equation}
such that $\kappa_e^{-1}$ is the grammage over which an electron of $E=1$~MeV loses its energy by ionization and plasma losses. Using $\kappa_e$, we may write
\begin{equation}\label{eq:dEdt}
  \frac{1}{\rho c}\frac{dE}{dt} = \frac{v_e}{c}\frac{dE}{dX}=1 g(E)\kappa_e ~\rm MeV,
\end{equation}
where $g(E)\approx1$ is a weak function of energy. For the relevant energy range (see below) of $\sim0.01$ to $\sim1$~MeV, we may approximate $g(E)=(E/1{\rm MeV})^{-0.15}$, reflecting a factor 2 change in the loss rate as the energy decreases from 1 to 0.01~MeV.

The time at which the ionization loss time is equal to the expansion time is given by $t_\varepsilon dE/dt=E$, or $\kappa_e\rho(t_\varepsilon) c t_\varepsilon g(E)=(E/1{\rm MeV})$. Since $t dE/dt\propto \rho t\propto t^{-2}$, the fraction of the energy of electrons produced at time $t\gg t_\varepsilon$, that is deposited in and heat the plasma, is $\approx(t/t_\varepsilon)^{-2}$. As pointed out in paper I, this is valid both for electrons that escape the plasma freely, in which case they spend a time $\sim r/c\propto t$ in the plasma, and for electrons that are confined to the plasma by magnetic fields. In the latter case, the electrons lose energy adiabatically, i.e contribute to the kinetic energy of the plasma by applying pressure against the expanding flow. Assuming that the confined energetic electrons behave as an ideal gas, we have $p\propto\rho^{4/3}$ and $E\propto\rho^{1/3}\propto t^{-1}$ for highly relativistic electrons, and $p\propto\rho^{5/3}$, $E\propto\rho^{2/3}\propto t^{-2}$ in the highly non-relativistic limit.
Both relativistic and non-relativistic electrons lose most of their energy to adiabatic expansion over time $t$, hence the fraction of their energy that is converted to thermal energy is $\approx(t/t_\varepsilon)^{-2}$.

For an ejecta mass $M$ and characteristic velocity $v_M$, neglecting the very weak dependence of $g$ on $E$ (in particular at $E\sim1$~MeV), we have
\begin{eqnarray}\label{eq:t_E}
  t_\varepsilon&=&f_\rho\left[\frac{c\kappa_e M}{4\pi E_{\rm MeV} v_{\rm M}^3}\right]^{1/2}
  \nonumber \\ &=& 2.5f_\rho
  \left(\frac{\kappa_e/E_{\rm MeV}}{1\rm cm^2/g} \frac{M}{0.01M_\odot}\right)^{1/2}v_{10}^{-3/2}\,{\rm d}.
\end{eqnarray}
Here, $E=1E_{\rm MeV}$~MeV, $v_M=10^{10}v_{10}{\rm cm/s}$ . $f_\rho$ is a dimensionless (order unity) coefficient that depends on the density distribution of the ejecta, defined such that the mass averaged value of the fraction of electron energy lost to ionization is given at late time by $(t/t_\varepsilon)^{-2}$.

Let us consider next the energy $E$ with which electrons are produced by beta-decay. In general, the decay-time is longer for lower $Q$-values, i.e. for lower energy release of the beta-decay. This behavior is due mainly to the larger phase space available for the emitted electron and neutrino. A rough lower limit to the decay time as a function of $Q$-value for $\beta^{-}$ decays is given by $\tau=10^4(Q/1\,{\rm MeV})^{-5}$\,s \citep[e.g.][]{Arnould2007PhR...450...97A}. The scatter in this relation is however very large- approximately two orders of magnitude at $Q=10$\,MeV, and six orders of magnitude at few MeV.
This spread is due to selection rules of the transitions, and it decreases with increasing $Q$-value as the number of final states increases. Thus, while we expect in general that the $Q$-value of the decays at the time scale in which we are interested would be $\sim1$\,MeV, one cannot adopt a simple relation between $Q$ (hence $E$) and $t$. A simple relation between $Q$ ($E$) and $t$ is unlikely to be valid also due to the fact that the decay of an isotope with a low $Q$-value and a long life time may produce an unstable isotope with a short life time and a high $Q$-value\footnote{For example, the 3~d, $0.5$\,MeV decay of $^{132}$Te produces $^{132}$I with 2~hr, 4~MeV decay.}.

The very large spread in decay times for $Q\sim$~few MeV, and the possible production of short lived isotopes by the decay of longer lived ones, imply that a time independent electron release energy $E$ may be more appropriate for our discussion. A more accurate treatment would require detailed calculations of the beta-decay chains. However, the accuracy of such calculations would at present be questionable, as experimental data for relevant heavy isotopes is partial at best, and theoretical estimates may deviate from experimental data by orders of magnitude \citep[see e.g.][]{Arnould2007PhR...450...97A}. For completeness, we nevertheless discuss in the derivation of the late time deposition rate (and bolometric luminosity), given in \S~\ref{sec:asym}, also in the case where $E$ is uniquely related to $t$. We find that the late time behavior is not sensitive to the time dependence of $E$.

Finally, let us consider the possible contribution of alpha-decays to the heating of the plasma. For low values of the initial electron fraction, $Y_e\lesssim0.1$, some nuclear network models, used in calculations of the isotopic evolution of merger ejecta, predict a significant, few tens of percent, contribution of alpha-decays to the radioactive energy release at $\sim100$~d \citep[e.g.][]{Barneskasen16}. The ionization loss rate of energetic massive particles (mass $\gg m_e$)
is given by \citep[e.g.][]{Longair92Book},
\begin{eqnarray}\label{eq:dEdX_M}
  \frac{dE}{dX}&=&\frac{4\pi z^2e^4}{m_e m_p v^2}\frac{Z}{A} \Big[\ln\left(\frac{2\gamma^2 m_e v^2}{\bar{I}}\right) -(v/c)^2\Big],
\end{eqnarray}
where $ze$, $v$ and $\gamma$ are the particle's charge, velocity and Lorentz factor, respectively.
alpha particles are expected to be produced with $\sim10$\,MeV kinetic energy. At this energy range, the ratio between the loss rate of alpha particles and electrons is approximately given by
\begin{equation}\label{eq:alpha_e}
  \frac{E_\alpha^{-1}\frac{dE_\alpha}{dt}}{E_e^{-1}\frac{dE_e}{dt}}\approx
  2(E_\alpha/10{\rm MeV})^{-3/2}(E_e/1{\rm MeV}).
\end{equation}
Thus, the release of a fraction of the energy in $\alpha$-particles instead of in electrons will lead to some increase in the energy deposition rate by particles produced at $\sim100$~d. This will, however, have only a minor effect on the energy deposition rate since, as we show in \S~\ref{sec:asym}, assuming that electrons are confined to the plasma, all electrons produced at $t\gtrsim t_\varepsilon$ (rather than only those produced at $t\sim100$~d) contribute to the ionization heating at $t\sim100$~d$\gg t_\varepsilon$.

\subsection{Asymptotic energy deposition for confined electrons}
\label{sec:asym}

We consider the ionization energy deposition by electrons, which are produced in beta-decays at a rate $\dot{n}(t)$ with initial energy $E_i(t)$.
Let us denote by $E(t_0,t)$ the time dependent energy of an electron produced at time $t_0$. The ionization heating rate at time $t$ is given by
\begin{equation}\label{eq:qd}
  \dot{\varepsilon}_d/1{\rm MeV}=\kappa_e\rho c\int dE \frac{dn(E,t)}{dE}g(E).
\end{equation}
Here, $dn/dE$ is the number of electrons per unit energy, and we have used Eq.~(\ref{eq:dEdt}) for the ionization loss rate. The differential electron number is given by
\begin{equation}\label{eq:dndE}
  \frac{dn(E,t)}{dE}=\dot{n}\left[t_0(E,t)\right]\frac{\partial t_0}{\partial E}.
\end{equation}
Note that $t_0(E,t)$ is the time at which an electron should be produced in order to have an energy $E$ at time $t$. We may therefore write
\begin{equation}\label{eq:qd_t}
  \dot{\varepsilon}_d/1{\rm MeV}=\kappa_e\rho c\int^t dt_0 \dot{n}(t_0) g[E(t_0,t)].
\end{equation}

In order to determine $\dot{\varepsilon}_d$, we need to determine $E(t_0,t)$. The evolution of the energy of the electrons is determined by
\begin{equation}\label{eq:Eeq}
  \frac{dE_{\rm MeV}}{dt}=-x\frac{E_{\rm MeV}}{t}-\kappa_e\rho c g(E).
\end{equation}
The first term on the r.h.s. accounts for adiabatic energy losses, and the second term accounts for ionization losses, following Eq.~(\ref{eq:dEdt}). As explained in \S~\ref{sec:late_rad_heat}, the term describing the adiabatic losses is obtained assuming that the high energy electrons behave as an ideal gas, in which case we have $x\approx1$ for highly relativistic electrons, and $x\approx2$ in the highly non-relativistic limit. This is of course only an approximate description. Moreover, the electrons are mildly relativistic, with $\gamma_e v_e/c\sim1$, and as they lose energy the value of $x$ that best describes the evolution is changing. We will solve below the evolution with a fixed value of $x$, and will show that due to the weak dependence of $g$ on $E$, the dependence of the late time evolution of the energy deposition on $x$ is small.

For simplicity, we will first discuss the case of time independent $E_i$, and will later provide the (straightforward) generalization to a time dependent $E_i$. For our analytic solutions, we will use a power-law approximation of $g(E)$,
\begin{equation}\label{eq:gE}
  g(E)\propto E^{-\omega_I},
\end{equation}
which holds with $\omega_I=0.15$ for $0.01<E_{\rm MeV}<1$
(see \S~\ref{sec:late_rad_heat} and Fig.~\ref{fig:dEdX}).
We show below that this is a good approximation by comparing the results of the analytic solutions to those obtained using the exact dependence of $g$ on $E$, as given by Eq.~(\ref{eq:dEdX}).

Measuring $t$ in units of $t_\varepsilon$, $\tilde{t}=t/t_\varepsilon$, and $E$ in units of $E_i$, $\epsilon=E/E_i$, Eq.~(\ref{eq:Eeq}) with the approximation of Eq.~(\ref{eq:gE}) becomes
\begin{equation}\label{eq:Eeq_DL}
  \frac{d\epsilon}{d\tilde{t}}=-x\frac{\epsilon}{\tilde{t}}-\frac{\epsilon^{-\omega_I}}{\tilde{t}^3}.
\end{equation}
The evolution of the energy of an electron produced at time $t_0$, $E(t_0,t)$, is determined by this equation with the initial condition $\epsilon(\tilde{t}_0,\tilde{t}_0)=1$.

Let us consider an electron produced at $\tilde{t_0}\gg1$ (i.e. $t_0\gg t_\varepsilon$). At production, the adiabatic losses are larger than the ionization losses by a factor $\tilde{t}_0^2$, and the energy evolution of the electron may be approximated using only the adiabatic loss term in Eq.~(\ref{eq:Eeq}), yielding
\begin{equation}\label{eq:highE}
  \epsilon=(\tilde{t}_0/\tilde{t})^x.
\end{equation}
As the electron loses energy, ionization losses may become significant. For the energy evolution of Eq.~(\ref{eq:highE}), the ratio of the ionization to adiabatic losses (second to first term in Eq.~(\ref{eq:Eeq_DL})) is $\epsilon^{-1-\omega_I}\tilde{t}^{-2}\propto \tilde{t}^{-2+x(1+\omega_I)}$.

Let us first consider the case $x\le2/(1+\omega_I)$. In this case, ionization losses never become important (for electrons produced at $\tilde{t}_0>1$). Eq.~(\ref{eq:highE}) holds at all times for all electrons produced at $\tilde{t}>1$, i.e. down to energy given by Eq.~(\ref{eq:highE}) with $\tilde{t}_0\approx1$,
\begin{equation}\label{eq:Em}
  \epsilon_m\approx \tilde{t}^{-x}.
\end{equation}
At time $\tilde{t}$, the electron distribution is strongly suppressed below $\epsilon_m(\tilde{t})$ since all electrons produced at $\tilde{t}<1$ lost most of their energy by ionization by $\tilde{t}<1$. We may therefore approximate
\begin{equation}\label{eq:qd_X1}
  \dot{\varepsilon}_d\propto t^{-3}\int_1^{\tilde{t}} d\tilde{t}_0\dot{n}(t_0) (\tilde{t}_0/\tilde{t})^{-x\omega_I}.
\end{equation}
Since the integral over $\dot{n}(t_0)$ must converge at large $t$, we have for $\tilde{t}\gg1$
\begin{equation}\label{eq:asym_X1}
   \frac{d\log\dot{\varepsilon}_d}{d\log t}=-3 +x\omega_I\quad {\rm for}\quad x\le2/(1+\omega_I).
\end{equation}
The asymptotic behavior is independent of the evolution of the radioactive energy release. This is due to the fact that, as can be seen from Eq.~(\ref{eq:qd_X1}), all electrons produced at $t_\varepsilon<t_0<t$ contribute to the ionization heating at time $t$, weakly weighted by $t_0^{-x\omega_I}$ (recall that $\omega_I\ll1$). Since $x\omega_I\ll1$, the convergence to the asymptotic behavior is determined by the convergence of $\int dt \dot{n}$.

\begin{figure}
\centerline{\includegraphics[width=8cm]{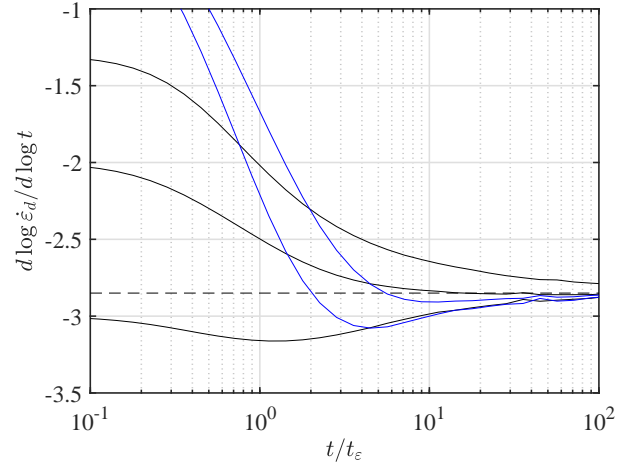}}
\caption{The evolution of the ionization heating rate, $\dot{\varepsilon_d}$, for various functional forms of the radioactive energy release rate, $\dot{\varepsilon}$. Black lines show the evolution for $\dot{\varepsilon}\propto t^{-\beta}$ with $\beta=1.3,2,3$ (from top to bottom), and blue lines for  $\dot{\varepsilon}\propto e^{-t/\tau}$ with $\tau/t_\varepsilon=0.5,1$. The dashed line is the asymptotic behavior given by Eq.~(\ref{eq:asym_X1}). For these calculations, $\omega_I=0.15$ and $x=1$.
\label{fig:asymR}}
\end{figure}

Fig.~\ref{fig:asymR} shows the evolution of $\dot{\varepsilon}_d$ as a function of time, obtained by numerically solving the complete implicit analytic solution of Eq.~(\ref{eq:Eeq_DL}) for $t_0(E,t)$ given in the appendix (and using it to numerically integrate Eq.~(\ref{eq:qd_t})). The solutions converge to the asymptotic behavior given by Eq.~(\ref{eq:asym_X1}) for a wide range of functional forms of $\dot{n}(t)$.

It is straightforward to generalize the above discussion for the case of a decreasing $E_i(t)\propto t^{-\omega_E}$. Normalizing the energy $E$ to $E_i(t_\varepsilon)$, Eq.~(\ref{eq:Eeq_DL}) remains unchanged, with initial conditions modified to $\epsilon(\tilde{t}_0,\tilde{t}_0)=\tilde{t}_0^{-\omega_E}$. Eq.~(\ref{eq:highE}) is replaced with $\epsilon=\tilde{t}_0^{-\omega_E}(\tilde{t}_0/\tilde{t})^x$, while Equation~(\ref{eq:Em}) remains unchanged (as long as $\omega_E<2/(1+\omega_I)$ so that at $t\gg t_\varepsilon$ the ionization-loss time of electrons produced at time $t$ is longer than $t$). The result of Eq.~(\ref{eq:asym_X1}) therefore also remains unchanged (as long as $\int dt_0 \dot{n}(t_0) t_0^{-(x-\omega_E)\omega_I}$ converges).

\begin{figure}
\centerline{\includegraphics[width=8cm]{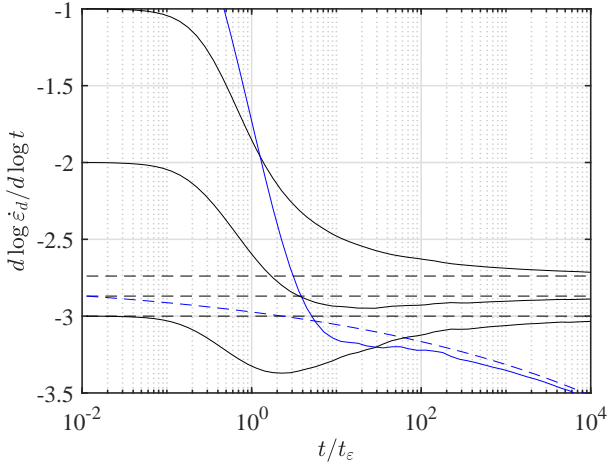}}
\caption{The evolution of the ionization heating rate, $\dot{\varepsilon_d}$, for various functional forms of the radioactive energy release rate, $\dot{\varepsilon}$, and highly non-relativistic electrons (which is not expected to be the case for beta-decay heating, see text). Black lines show the evolution for $\dot{\varepsilon}\propto t^{-\beta}$ with $\beta=1,2,3$ (from top to bottom), and blue lines for  $\dot{\varepsilon}\propto e^{-t/\tau}$ with $\tau/t_\varepsilon=1$. The dashed lines are the asymptotic behavior given by Eqs.~(\ref{eq:asym_X2}) and~(\ref{eq:asym_X2exp}). For these calculations, $\omega_I=0.15$ and $x=2$.
\label{fig:asym}}
\end{figure}

Noting that $\omega_I\ll1$, the regime $x>2/(1+\omega_I)\approx 2$ is not relevant for our discussion, since $x=2$ is obtained only in the highly non-relativistic limit, while in our case the beta-decay electrons remain mildly relativistic over the relevant time. The energy of electrons produced at $\tilde{t}\sim1$
($t\sim t_\varepsilon$) with $E\sim 1$\,MeV, drops to $\sim0.1$\,MeV with $v_e/c=0.6$ at $t=10t_\varepsilon$, and to $\sim0.01$\,MeV with $v_e/c=0.2$ at $t=100t_\varepsilon$. Nevertheless, we give below the results for this case as well, both for completeness and for demonstrating that the results are not strongly modified also for $x=2$.

For $x>2/(1+\omega_I)$, as the electron cools adiabatically following Eq.~(\ref{eq:highE}), it eventually reaches an energy at which ionization losses become important. Comparing the two loss terms of Equation~(\ref{eq:Eeq_DL}), we find that at time $\tilde{t}$ ionization losses dominate below
\begin{equation}\label{eq:Ec}
  \epsilon_c\approx \tilde{t}^{-2/(1+\omega_I)}.
\end{equation}
Using Eq.~(\ref{eq:highE}), we find that electrons reaching this energy at time $\tilde{t}$ were produced at,
\begin{equation}\label{eq:t_c}
  \tilde{t}_c=\tilde{t}_0(\epsilon_c,\tilde{t})=\left(\epsilon_c \tilde{t}^x\right)^{1/(x-\omega_E)}\approx \tilde{t}^{\frac{(1+\omega_I)x-2}{(x-\omega_E)(1+\omega_I)}}.
\end{equation}
At time $\tilde{t}$, the electron distribution is strongly suppressed below $\epsilon_c(\tilde{t})$ since all electrons produced at $\tilde{t}<\tilde{t}_c$ lost most of their energy by ionization. We may therefore approximate
\begin{equation}\label{eq:qd_Xh}
  \dot{\varepsilon}_d\propto t^{-3}\int_{\tilde{t}_c}^{\tilde{t}} d\tilde{t}_0\dot{n}(t_0) (\tilde{t}_0/\tilde{t})^{-x\omega_I}.
\end{equation}
Since $\tilde{t}_c$ diverges as $\tilde{t}$ diverges, and since $\dot{n}(t)$ drops faster than $1/t$, the integral is dominated by the contribution from $t_0\sim t_c$. In this case, the asymptotic behavior does depend on the temporal evolution of $\dot{n}$ and of $E_i$, as the ionization heating is determined at time $t$ by electrons produced at time $t_0\gtrsim t_c(t)$.

For a power-law  behavior, $\dot{n}\propto t^{-(\beta-\omega_E)}$ and $E_i\propto t^{-\omega_E}$, and hence $\dot{\varepsilon}\propto t^{-\beta}$, we find
\begin{equation}\label{eq:asym_Xh}
  \frac{d\log\dot{\varepsilon}_d}{d\log t}=-3+\frac{2\omega_I}{1+\omega_I}-\frac{[(1+\omega_I)x-2](\beta-\omega_E-1)}{(1+\omega_I)(x-\omega_E)}.
\end{equation}
This is the result obtained by \citet{2018KB_dEdX} (see their Eq.~(26)). For $x=2$,
\begin{equation}\label{eq:asym_X2}
  \frac{d\log\dot{\varepsilon}_d}{d\log t}=-3+\frac{2\omega_I}{1+\omega_I}\frac{(3-\beta)}{(2-\omega_E)}.
\end{equation}
For $x=2$ and an exponential $\dot{\varepsilon}\propto\dot{n}\propto e^{-t/\tau}$, with time independent $E_i$, we have
\begin{equation}\label{eq:asym_X2exp}
    \frac{d\log\dot{\varepsilon}_d}{d\log t}=-3+\frac{2\omega_I}{1+\omega_I}-f(\omega_I)\frac{t_\varepsilon}{\tau}
    \left(\frac{t}{t_\varepsilon}\right)^{\frac{\omega_I}{1+\omega_I}}.
\end{equation}
Since $\omega_I\ll1$, we may approximate for the times of interest
\begin{equation}\label{eq:asym_X2exp_app}
    \frac{d\log\dot{\varepsilon}_d}{d\log t}\simeq-3+\frac{2\omega_I}{1+\omega_I}-f(\omega_I)\frac{t_\varepsilon}{\tau}.
\end{equation}
In order to determine $f(\omega_I)$ an exact relation between $\tilde{t}_c$ and $\tilde{t}$ is required. Using Eq.~(\ref{eq:t0l}) of the Appendix we find
\begin{equation}\label{eq:f_w}
  f(\omega_I)=\frac{1}{2}\left(\frac{2\omega_I}{1+\omega_I}\right)^{1-\frac{1}{2(1+\omega_I)}}.
\end{equation}
Fig.~\ref{fig:asym} shows the evolution of $\dot{\varepsilon}_d$ as a function of time, for $x=2$ and a wide range of functional forms of $\dot{n}(t)$. The solutions converge to the asymptotic behavior given by Eqs.~(\ref{eq:asym_X2}) and ~(\ref{eq:asym_X2exp}).

\begin{figure}
\centerline{\includegraphics[width=8cm]{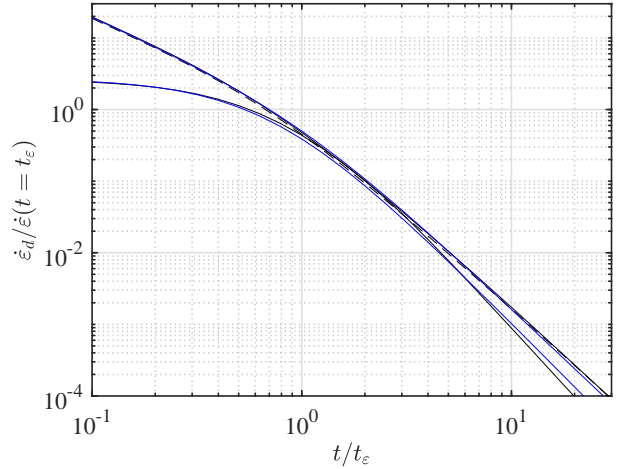}}
\caption{Solid black curves give $\dot{\varepsilon}_d$ obtained for a power-law, $\dot{\varepsilon}\propto t^{-1.3}$, and an exponential, $\dot{\varepsilon}\propto e^{-t/t_\varepsilon}$, radioactive energy release rates, approximating the energy dependence of the electron energy loss rate by $g(E)\propto E^{-0.15}$. Blue curves are interpolations between the asymptotic $t\ll t_\varepsilon$ behavior, $\dot{\varepsilon}_{d0}=\dot{\varepsilon}=t^{-1.3},\, t^0$, and the asymptotic $t\gg t_\varepsilon$ behavior, $\dot{\varepsilon}_{d\infty}$, given in Eqs.~(\ref{eq:asym_X2}) and~(\ref{eq:asym_X2exp_app})- $(\dot{\varepsilon}_{d0}^m+\dot{\varepsilon}_{d\infty}^m)^{1/m}$ with $m=-1,-2$ for the power-law and exponential cases. The dashed curve shows the solution obtained using the full $g(E)$ dependence for Pb given by Eq.~(\ref{eq:dEdX}) (with $E_i=1$~MeV).
\label{fig:complete}}
\end{figure}

The convergence of the solutions to the asymptotic behavior appears slow, both for $x=1$ (Fig.~\ref{fig:asymR}) and for $x=2$. In Fig.~\ref{fig:complete} we show, however, that simple interpolations between the $t\gg t_\varepsilon$ asymptotic behavior and the $t\ll t_\varepsilon$ asymptotic behavior, where $\dot{\varepsilon}_d=\dot{\varepsilon}$, provide excellent approximations to the complete solutions. This figure also compares $\dot{\varepsilon}_d$ obtained using the approximation $g(E)\propto E^{-0.15}$ and that obtained using the exact $g(E)$ given by Eq.~(\ref{eq:dEdX}). The results are nearly identical.

\subsection{Summary and analytic light curves}
\label{sec:analytic_lc}

The ionization energy-loss rate of electrons is nearly independent of their energy $E$ and of the plasma composition (see Fig.~\ref{fig:dEdX}). It is given by Eq.~(\ref{eq:dEdt}) with an effective ionization "opacity", $\kappa_e\approx 1$\,cm$^{2}$g$^{-1}$, and a weak energy dependence given by $g(E)$. For the relevant energy range of $\sim0.01$ to $\sim1$\,MeV, we may approximate $g(E)\propto E^{-\omega_I}$ with $\omega_I=0.15$, reflecting a factor 2 change in the loss rate as the energy decreases from 1 to 0.01~MeV.

For an ejecta mass $M$ and characteristic velocity $v_M$, neglecting the very weak dependence of $g$ on $E$ (in particular at $E\sim1$~MeV), we have (Paper I)
\begin{eqnarray}\label{eq:t_Es}
  t_\varepsilon&=&f_\rho\left[\frac{c\kappa_e M}{4\pi E_{\rm MeV} v_{\rm M}^3}\right]^{1/2}
  \nonumber \\ &=& 2.5f_\rho
  \left(\frac{\kappa_e/E_{\rm MeV}}{1\rm cm^2/g} \frac{M}{0.01M_\odot}\right)^{1/2}v_{10}^{-3/2}\,{\rm d}.
\end{eqnarray}
Here, $E=1E_{\rm MeV}$~MeV, $v_M=10^{10}v_{10}{\rm cm/s}$ . $f_\rho$ is a dimensionless (order unity) coefficient that depends on the density distribution of the ejecta, defined such that the mass averaged value of the fraction of electron energy lost to ionization is given at late time by $(t/t_\varepsilon)^{-2}$. In Paper 1 we have considered an ejecta with a power-law dependence of mass on velocity, $dm/dv\propto v^{-(\alpha+1)/\alpha}$ for $v\ge v_M$, for which $f_\rho=[\alpha(2+3\alpha)]^{-1/2}$ (for GW\,170817 observations imply $\alpha\approx0.6$ and hence $f_\rho\approx0.7$).

Eq.~(\ref{eq:t_Es}) generalizes Eq.~(10) of Paper I, where $E=1$~MeV was assumed. As discussed in \S~\ref{sec:late_rad_heat}, while in general the decay-time is longer for lower energy ($Q$-value) decays, with a rough lower limit to the decay time as a function of $Q$-value given by $\tau=10^4(Q/1\,{\rm MeV})^{-5}$\,s \citep[e.g.][]{Arnould2007PhR...450...97A}, one cannot adopt a simple relation between $Q$ (hence $E$) and $t$. The scatter in this relation is very large- approximately two orders of magnitude at $Q=10$\,MeV, and six orders of magnitude at few MeV, and the decay of an isotope with a low $Q$-value and a long life time may produce an unstable isotope with a short life time and a high $Q$-value. This suggests that a time independent electron release energy $E\sim1$~MeV may be more appropriate for the times under consideration.

If the electrons freely escape the plasma, then at $t>t_\varepsilon$ we simply have $\dot{\varepsilon}_d\approx\dot{\varepsilon}(t/t_\varepsilon)^{-2}$. For confined electrons, we have shown that
\begin{equation}\label{eq:asym_PL}
  \dot{\varepsilon}_d\propto t^{-3+x\omega_I},
\end{equation}
independent of the evolution of the radioactive energy release, as well as of the possible evolution of the energy $E$ with which electrons are produced. Here $x$ is the energy decay exponent describing adiabatic expansion losses, $dE/dt=-x E/t$. This description of adiabatic losses holds if the high energy electrons behave as an ideal gas, in which case we have $x\approx1$ for highly relativistic electrons and $x\approx2$ for highly non-relativistic electrons. This is obviously a rough approximation. However, since $\omega_I\ll1$, the result is not sensitive to the exact value of $x$, and for the mildly relativistic electrons, $\gamma_e v_e/c\sim1$, in which we are interested we may use $x= 1.5$, which implies $-3+x\omega_I\simeq-2.8$.

The convergence of the solutions to the asymptotic behavior derived in \S~\ref{sec:asym} appears slow, see Fig.~\ref{fig:asymR} and Fig.~\ref{fig:asym}. In Fig.~\ref{fig:complete} we show, however, that simple interpolations between the $t\gg t_\varepsilon$ asymptotic behavior and the $t\ll t_\varepsilon$ asymptotic behavior, where $\dot{\varepsilon}_d=\dot{\varepsilon}$, provide excellent approximations to the complete solutions. This figure also shows that $\dot{\varepsilon}_d$ obtained using the approximation $g(E)\propto E^{-0.15}$ and that obtained using the exact $g(E)$ given by Eq.~(\ref{eq:dEdX}) are nearly identical.

At times later than $t_M$, defined as the time at which the photon diffusion time out of the ejecta equals the expansion time $t$, the bolometric luminosity is given by $L= \dot{\varepsilon}_d$. For $t_M<t_\varepsilon$, as is the case for GW\,170817 (see Paper I), we thus have for $t>t_\varepsilon$ and $\dot{\varepsilon}\propto t^{-\beta}$
\begin{equation}\label{eq:Lb}
  L \propto \left\{
              \begin{array}{ll}
                (t/t_\varepsilon)^{-\beta-2}, & \hbox{free electron escape;} \\
                (t/t_\varepsilon)^{-3+x\omega_I\simeq-2.8}, & \hbox{electron confinement.}
              \end{array}
            \right.
\end{equation}
This corrects equations~(12, A23) of Paper I to include the effects of the accumulation of electrons and of the energy dependence of the ionization loss rate.

Our results differ from those of \citet{2018KB_dEdX} for two reasons. First, they approximate the energy dependence of the ionization loss rate as $g(E)\propto E^{-\omega_I}$ with $\omega_I=0.5$ (they do mention that the value may be slightly smaller, but adopt $\omega_I=0.5$ as the "default" value used to determine their predictions for a power-law energy deposition, and as the only value for which results are derived for an exponential deposition; this value was later adopted by \citet{2018Kasliwal_IR} for the analysis of the late time observations of GW\,170817). Choosing $\omega_I=0.5$ was motivated by noting that $dE/dt\propto v_e dE/dX\propto v_e^{-1}\log(E)$ (see Eq.(\ref{eq:dEdX})), neglecting the $\log(E)$ dependence and using the non-relativistic approximation $v_e\propto E^{1/2}$. These approximations are not accurate for the energy range of interest, producing a much steeper energy dependence than that given by Eq.~(\ref{eq:dEdX}), as can be clearly seen also from examining Fig.~(\ref{fig:dEdX}). This large value of $\omega_I$ leads to the prediction of a much shallower decline of $L= \dot{\varepsilon}_d$. Second, they describe adiabatic losses using $x=2$, which is appropriate only in the highly non relativistic limit, and therefore not applicable for the current discussion. As explained in \S~\ref{sec:asym}, the result of Eq.~(\ref{eq:asym_PL}) holds for $x<2/(1+\omega_I)$, while for $x>2/(1+\omega_I)$ the asymptotic behavior is given by Eq.~(\ref{eq:asym_Xh}), which reduces to Eq.~(\ref{eq:asym_X2}) for $x=2$ \citep[which is the result obtained by][]{2018KB_dEdX}. It should be noted that for $\omega_I\ll1$ the asymptotic decay index is $\approx-3$, independent of $x$ and $\beta$. Hence, using Eq.~(\ref{eq:asym_X2}) with $\omega_I=0.15$ would yield a late time decay index which is close to that of Eq.~(\ref{eq:Lb}), in the range of $-2.8$ to $-3$ for $1.1<\beta<3$ and $\omega_E=0$.

\section{The case of GW\,170817}
\label{sec:GW170817}

In this section, we discuss the implications of the late-time observations of GW\,170817. The observations are discussed in \S~\ref{sec:obs}, and the analysis and implications is discussed in \S~\ref{sec:obs_anal}.

\subsection{Observations}
\label{sec:obs}

For the early time observations ($t<20$\,days) we adopt the bolometric light curve of Paper I. This bolometric light curve was estimated by polynomialy interpolating all the available observations in all the bands, and integrating over wavelength (see details in Paper I). We note that using bolometric light curves from other sources does not change our conclusions.
This is demonstrated by using also the bolometric light curve presented in \citet{2017Kasliwal} (see Figure~\ref{fig:Kasliwal}, \S~\ref{sec:obs_anal}).

The main sources of the late time visible-light and IR
observations ($t>20$\,days) are \cite{2018Kasliwal_IR} and \cite{Lamb2019}. At late time, of order tens of days, synchrotron emission from the collisionless shock driven by the ejecta into the surrounding medium, which is responsible for the observed X-ray and radio emission, may contribute a significant fraction of the flux observed at the UV-IR bands. The removal of this component from the optical observations is discussed in \S~\ref{sec:synch}.

On tens of days time scale, the synchrtron luminosity increases with time while the radioactive kilonova emission decreases with time. We show that while the kilonova emission dominates at the time of the first Spitzer observations, $t=43$\,day, by $t=74$\,days the synchrotron contribution is estimated as $\approx25$\% with large uncertainty (due to uncertainty in modelling the synchrotron emission, which may deviate from a simple interpolation between radio and X-ray measurements). At still later times (i.e., {\it HST} observations), the observed flux is consistent with being dominated by the synchrotron component, rendering credible estimates of the kilonova emission at these late times impossible.

\begin{deluxetable*}{llllll}
\tablewidth{0pt}
\tablecaption{Late time observations of GW\,170817}
\tablehead{
\colhead{Time}          &
\colhead{Frequency}     &
\colhead{$L_{\nu}$}     &
\colhead{$\Delta{L_{\nu}}$} &
\colhead{Instrument}    &
\colhead{Reference}   \\
\colhead{(day)}        &
\colhead{(Hz)}       &
\colhead{(erg\,s$^{-1}$\,Hz$^{-1}$)}       &
\colhead{(erg\,s$^{-1}$\,Hz$^{-1}$)}       &
\colhead{}     &
\colhead{}
}
\startdata
   9.20  &  $7.25\times10^{16}$ &  $1.65\times10^{21}$ & $8.42\times10^{20}$ &  Chandra      & Nynka et al. 2018 \\
   9.21  &  $7.25\times10^{16}$ &  $1.12\times10^{21}$ & $4.81\times10^{20}$ &  Chandra      & Margutti et al. 2018 \\
  10.37  &  $6.20\times10^{9}$  &  $1.49\times10^{25}$ & $4.98\times10^{24}$ &  VLA          & Hallinan et al. 2017 \\
  15.39  &  $7.25\times10^{16}$ &  $1.53\times10^{21}$ & $4.66\times10^{20}$ &  Chandra      & Margutti et al. 2018 \\
  15.60  &  $7.25\times10^{16}$ &  $7.41\times10^{21}$ & $2.67\times10^{21}$ &  Chandra      & Nynka et al. 2018
\enddata
\tablecomments{A collection of all the X-ray and radio data for GW\,170817, as well
as the Spitzer-IR and {\it HST} visible-light observations. The X-ray effective frequency was calculated using Equation~(\ref{eq:EffE}), and the reported $\Gamma$. When available, we used measurements based on combining several contiguous days. Here we present the first five lines, while the full table is available at {\tt http://euler1.weizmann.ac.il/papers/GW170817late/Table\_Lnu\_wHead.txt}.
Measurments are adopted from \cite{Nynka2018}, \cite{Margutti2018a}, \cite{Margutti2018b}, \cite{Hallinan2017}, \cite{2018Kasliwal_IR}, \cite{Lamb2019}, \cite{Alexander2018}, \cite{Dobie2018}, \cite{Resmi2018}, \cite{Mooley2018a}, \cite{2Mooley2018b}, \cite{Troja2018}, and \cite{Haggard2018}.}
\label{tab:Obs}
\end{deluxetable*}

\subsubsection{The synchrotron component}
\label{sec:synch}

Table~\ref{tab:Obs} lists all available observations taken at $t>20$\,days, along with their references.
Since the X-ray observations are usually obtained
over a very wide band, we calculated the specific luminosity
using the best fit photon power-law index $\Gamma$, defining
the specific luminosity at the lower-limit of the band ($\nu_{1}$) as
\begin{equation}
    L_{\nu}(\nu_{1}) = (2-\Gamma) \frac{\nu_{1}^{1-\Gamma}  }{\nu_{2}^{2-\Gamma} - \nu_{1}^{2-\Gamma}} L .
    \label{eq:EffE}
\end{equation}
Here $nu_{2}$ is the band frequency upper limit.

In order to be able to remove the synchrotron component
from the optical observations, we fitted the X-ray and radio
specific luminosity, $L_{\nu}$ given in Table~\ref{tab:Obs},
with a physically motivated simple model of the form
\begin{equation}
    L_{\nu} = A t^{a} \nu^{b}.
\label{eq:SynModel}
\end{equation}
Here $\nu$ is the frequency, and $A$, $a$ and $b$ are
free parameters, chosen independently for different time intervals. The best fitted $A$, $a$, and $b$ in some selected time bins are listed in Table~\ref{tab:Fit}. The time bins in which we performed the fit
were selected by trial and error to represent periods
in which the observations are consistent with this simple model.

As can be seen from Table~\ref{tab:Fit}, the ratio of X-ray and radio specific luminosities is independent of time, with $b\simeq-0.57$. This is consistent with synchrotron emission from a power-law distribution of electrons, $dN/dE\propto E^{-p}$, which do not radiate a significant fraction of their energy over the observed time. In this case $p=1-2b\cong2.14$. A power-law distribution with $p\gtrsim2$ is expected for collisionless shock acceleration \citep[e.g.][]{Marcowith2016RPPh...79d6901M}, supporting the slow-cooling synchrotron emission interpretation.

\begin{deluxetable*}{lllrlllll}
\tablecolumns{9}
\tablewidth{0pt}
\tablecaption{Fitted parameters for the synchrotron model}
\tablehead{
\colhead{mean time}   &
\colhead{time range}  &
\colhead{$\log_{10}[A/({\rm erg\,s^{-1}Hz^{-1}})]$}           &
\colhead{$a$}     &
\colhead{$b$}    &
\colhead{$\chi^{2}$/dof}  &
\colhead{RMS}          &
\colhead{$N_{\rm X}$}  &
\colhead{$N_{\rm R}$}  \\
\colhead{(day)}     &
\colhead{(day)}     &
\colhead{}          &
\colhead{}          &
\colhead{}          &
\colhead{}          &
\colhead{}          &
\colhead{}          &
\colhead{}
}
\startdata
 17.1 &   9-- 30 & $ 29.555\pm  0.782$ & $ 1.09\pm 0.51$ & $-0.56\pm 0.02$ &  10.9/13 & 0.204 &  5 & 11 \\
 94.8 &  20--150 & $ 29.823\pm  0.203$ & $ 0.75\pm 0.09$ & $-0.55\pm 0.01$ &  16.7/31 & 0.106 &  2 & 32 \\
132.7 & 100--180 & $ 30.674\pm  0.576$ & $ 0.37\pm 0.26$ & $-0.55\pm 0.01$ &  21.1/29 & 0.108 &  3 & 29 \\
171.3 & 150--200 & $ 31.031\pm  1.417$ & $ 0.30\pm 0.62$ & $-0.57\pm 0.01$ &   4.9/14 & 0.074 &  1 & 16 \\
225.8 & 180--275 & $ 36.210\pm  0.579$ & $-1.98\pm 0.25$ & $-0.58\pm 0.01$ &   3.2/18 & 0.060 &  1 & 20 \\
256.7 & 210--400 & $ 34.491\pm  0.483$ & $-1.25\pm 0.21$ & $-0.58\pm 0.01$ &   4.8/17 & 0.058 &  2 & 18
\enddata
\tablecomments{Best fit power-law in frequency and time to the X-ray and radio data (Eq.~\ref{eq:SynModel}) in some selected date ranges. $N_{\rm X}$ and $N_{\rm R}$ are the numbers of X-ray and radio points used in the fit, respectively.}
\label{tab:Fit}
\end{deluxetable*}

The single power-law fit of Equation~\eqref{eq:SynModel} is valid assuming that the "cooling frequency", $\nu_c$, above which radiation is produced by electrons that lose most of their energy over the observed time, is above the X-ray frequency, $\nu_X$. For $\nu_c<\nu_X$, a broken power-law should be fitted, with $L_\nu\propto \nu^{-p/2}$ for $\nu>\nu_c$. Such a fit would yield a smaller value of $p$, and a larger estimate of the flux in the optical/IR bands. We note that $\nu_c<\nu_X$ is unlikely since we generally expect $\nu_c$ to decrease with time, which will lead to variability in $b$ over time for $\nu_c<\nu_X$. This is not observed (as can be seen in Table \ref{tab:Fit}). Moreover, for a smaller value of $p$ the predicted optical flux will exceed the flux measured by {\it HST} (see Fig.~\ref{fig:BolLC_withIR}).
We also note that the X-ray spectral power-law index ($\Gamma\cong1.6$) is consistent with the inferred value of $p$.

Table~\ref{tab:Ratio} presents the ratio between the observed IR and visible-light flux and the synchrotron flux interpolated into the optical wavebands. The error term includes only the reported relative photometric errors without any uncertainty related to the interpolation of the synchrotron flux. At the first Spitzer observations ($t=43$\,day), the kilonova emission dominates the observed flux. However, by $t=74$\,days the synchrotron is contributing about 25\% of the observed flux, and the exact fraction is highly uncertain due to the uncertainty in the interpolation of the synchrotron emission.
At later epochs (i.e., {\it HST} observations), the observed flux is consistent with being dominated by the synchrotron emission.

\begin{deluxetable}{llll}
\tablecolumns{4}
\tablewidth{0pt}
\tablecaption{Ratio between observed flux and interpolated synchrotron flux}
\tablehead{
\colhead{time}      &
\colhead{Band}      &
\colhead{Obs/pred.}     &
\colhead{Error}    \\
\colhead{(day)}     &
\colhead{}          &
\colhead{}          &
\colhead{}
}
\startdata
43     & 4.5$\mu$m  &    39   & 0.05 \\
74     & 4.5$\mu$m  &    4.1  & 0.2  \\
\hline
172.12 & F810W      & $\sim0.84$  & 0.2  \\
172.19 & F606W      & $\sim0.5$    & 0.2  \\
296.74 & F606W      & 0.7    & 0.2  \\
326.61 & F606W      & 0.6    & 0.2  \\
361.7 & F606W       & 0.4    & 0.4
\enddata
\label{tab:Ratio}
\end{deluxetable}

\subsection{Implications of the late time kilonova emission}
\label{sec:obs_anal}

\begin{figure}
\centerline{\includegraphics[width=8cm]{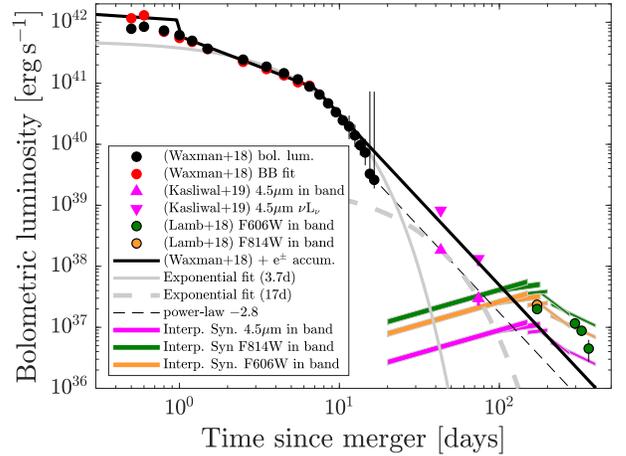}}
\caption{The kilonova bolometric light curve based on the Paper I integral of spectral energy distribution reduction (black circles) and black body fit (red circles, up to 7~d, beyond which the emission is highly non thermal), and the late time IR and optical observations. At early time, $t<0.8$~d, the temperature is high and a significant fraction of the flux may reside outside the observed bands, hence the black circles represent a lower limit to $L$. The upper-pointed tip triangles show the integrated luminosity in the Spitzer 4.5$\mu$m band \citep{2018Kasliwal_IR}, while the lower-tip triangles show $\nu L_{\nu}$. The yellow and green circles (with black edge) show the {\it HST} measurements \citep{Lamb2019}, corrected for the foreground galactic extinction ($E_{B-V}=0.11$\,mag). Luminosities are calculated for a source distance of 40\,Mpc. The black line shows the simple analytic model of Paper I, with $L\propto t^{-2.8}$ at $t>t_\varepsilon$ (instead of $L\propto t^{-3}$ in paper I, to correct for the accumulation of low energy electrons). The black line is plotted using the parameters adopted in paper~I (an exact best fit was not attempted due to parameter degeneracy). We note that this model also agrees very well with the observed effective temperature (see paper~I). The solid gray line shows the best fitted exponential model for the bolometric light curve at $3\,{\rm d}<t<17$\,d, with exponential time scale of $3.6\pm0.2$~d, while the dashed gray line shows an exponential fit to the IR data with a time scale of 17~d (see discussion in \S~\ref{sec:discussion}). The dashed line shows $L\propto t^{-2.8}$ with the minimum normalization required in order not to contradict the IR observations. It intersects the 3.7~d exponential fit at about three weeks.
The magenta, yellow and green lines show the synchrotron fit (Eq.~(\ref{eq:SynModel}), Table~\ref{tab:Fit}) interpolated to the 4.5$\mu$m, F814W, and F606W bands, with fit rms given by the line width. The synchrotron emission is dominated by the kilonova radioactive emission at $t<100$~d, but likely dominates the flux at later time.
\label{fig:BolLC_withIR}}
\end{figure}

\begin{figure}
\centerline{\includegraphics[width=8cm]{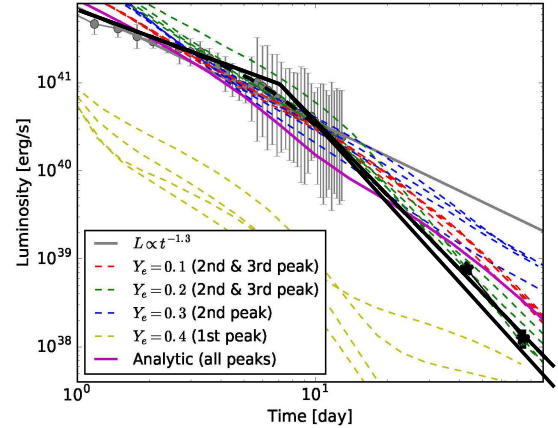}}
\caption{The analytic model of Paper I (solid black curves)  overlaid on the bolometric light curve of \citet{2018Kasliwal_IR} and the late time Spitzer observations as derived there \citep[Adopted from Figure 2 of][]{2018Kasliwal_IR}. At late time we show both the $t^{-3}$ behavior of Paper I, and the $t^{-2.8}$ behavior of the improved model, Eq.~(\ref{eq:Lb}), including the effects of the accumulation of electrons and the energy dependence of ionization losses. The dashed black curve is a simple interpolation between the two power-laws at $t<7$~d and $t>7$~d. The colored lines show the results of the light curve calculations of \citet{2018Kasliwal_IR}. Our calculations differ from theirs in the treatment of electron energy deposition and in the assumed ejecta velocity distribution (see \S~\ref{sec:Intro} and Paper I). \label{fig:Kasliwal}}
\end{figure}

Figure~\ref{fig:BolLC_withIR} presents the kilonova bolometric light curve as well as the late time Spitzer observations and some fitted models. The black circles represent the bolometric light curve derived in Paper I. The \cite{2018Kasliwal_IR} IR observations are shown as magenta triangles, where the upper-pointed tip triangles show the integrated luminosity in the Spitzer 4.5$\mu$m band (i.e., lower limit on the bolometric
luminosity), while the lower-tip triangles show $\nu L_{\nu}$. The black line shows the simple analytic model of Paper I, for which $t_\varepsilon=7$~d, with $L\propto t^{-2.8}$ at $t>t_\varepsilon$ instead of $L\propto t^{-3}$ in paper I, to correct for the accumulation of low energy electrons. The gray line shows the best fitted exponential model for $t>3$\,days with exponential time scale of $3.6\pm0.2$\,days. Figure~\ref{fig:Kasliwal} shows the analytic model of Paper I
overlaid on the bolometric light curve, including the late time Spitzer points, as derived in \citet{2018Kasliwal_IR}.

Prior to discussing the implication of the late time observations, the following point should be noted. On a time scale of tens of days the plasma is optically thin, far from
Local Thermodynamic Equilibrium (LTE), and the emission is far from thermal. The luminosity will be produced at wavebands where atomic transitions with sufficiently strong oscillator strengths and sufficiently large population of excited states exit. Without a detailed understanding, which is currently lacking, of the oscillator strengths and excitation cross sections of the relevant atoms, the luminosity measured at a given band provides only a lower limit to the bolometric luminosity, rather than a robust estimate of it. Thus, while models that predict a bolometric luminosity lower than that observed within a single band can be ruled out, models predicting larger bolometric luminosity would be consistent with the data.

The late time IR observations are consistent with the analytic model of Paper I, for which $t_\varepsilon\approx 7$\,d. In models with $t_\varepsilon\gg 7$\,d, the steepening of the bolometric light curve must be explained as due to a suppression of the radioactive energy release rate, $\dot{\varepsilon}$. As pointed out in Paper I, and shown in Fig.~\ref{fig:BolLC_withIR}, the 3-17~d data are consistent with an exponential suppression, $\dot{\varepsilon}\propto e^{-t/\tau}$ with $\tau\approx 3.6$\,d. Such a suppression could occur if the radioactive energy release is dominated by the decay of isotopes with the appropriate life time. In this case, $t_\varepsilon\lesssim 20$\,d would be required, with $L\propto t^{-2.8}$ at $t>20$\,d, in order for the bolometric luminosity not to fall below the observed IR luminosity (see Fig.~\ref{fig:BolLC_withIR}). We note that if the bolometric luminosity is larger than the observed luminosity in the 4.5$\,\mu$m band, then the upper limit on $t_\varepsilon$ is lower.

\section{Discussion}
\label{sec:discussion}

The inefficient ionization energy loss of beta-decay electrons at $t>t_\varepsilon$ leads to a steepening of the bolometric light curve. $t_\varepsilon$ is given by Eq.~(\ref{eq:t_Es}), where $E$ is the characteristic energy with which beta decay electrons are released, and $\kappa_e$ is defined in Eq.~(\ref{eq:kappa_e}). Its value is nearly independent of the composition of the ejecta, $\kappa_e\approx1{\rm cm^2/g}$.

For the case where beta-decay electrons are confined to the expanding ejecta, the asymptotic behavior is given by Eq.~(\ref{eq:asym_X1}), $d\log\dot{\varepsilon}_d/d\log t=-3+x\omega_I$, where $\omega_I\approx0.15$ describes the  dependence of the ionization loss rate on the electron energy, $dE/dt\propto E^{-\omega_I}$ (see \S~\ref{sec:late_rad_heat}) and $x$ characterizes the adiabatic energy loss rate, $dE/dt=-xE/t$ (with $x=1$ for highly relativistic electrons and $x=2$ in the highly non-relativistic limit). For the mildly relativistic electrons of interest, we have $d\log\dot{\varepsilon}_d/d\log t\simeq-2.8$. The asymptotic behavior is independent of the time dependence of the radioactive energy release rate, and of the possible temporal dependence of the kinetic energy with which beta decay electrons are produced. This implies that the late time bolometric light curve does not provide strong constraints on the composition of the ejecta.

The late time IR observations of GW\,170817 are consistent with the analytic model of Paper I, for which $t_\varepsilon\approx 7$~d, see Figs.~(\ref{fig:BolLC_withIR}) and~(\ref{fig:Kasliwal}). Models with larger $t_\varepsilon$ values may also be constructed to explain the data. In such models the steepening of the bolometric light curve at $\approx7$~d may be explained as due to a suppression of the radioactive energy release rate, $\dot{\varepsilon}$, which may be obtained if the radioactive energy release is dominated (at $\sim7$~d) by the decay of isotopes with $\approx4$~d life time (see Fig.~\ref{fig:BolLC_withIR}). The late time IR measurements may be consistent with such models if $t_\varepsilon\lesssim20$~d, or, for $t_\varepsilon>20$~d, if isotopes with $\approx 17$~d life time and appropriate abundance dominate the energy release at late time (see Fig.~\ref{fig:BolLC_withIR}). For $t_\varepsilon\gg7$~d, the agreement of the light curve with the predicted steepening due to adiabatic losses at 7~d would be a coincidence.

At late time, the plasma is optically thin and far from LTE, and the luminosity is produced at wavebands where atomic transitions with sufficiently strong oscillator strengths and sufficiently large population of atoms in excited states exist. Inferring the total bolometirc luminosity from single band observations is therefore difficult. The fact that a considerable part of the emission at $t\sim50$\,d is at the 4.5$\mu$m band (and not detected in the 3.6$\mu$m band), provides an important handle on the composition of the plasma. Its interpretation requires a detailed understanding, which is currently lacking, of the oscillator strengths, the excitation cross sections of the relevant atoms, and the energy distribution of the beta-decay electrons.

Let us finally comment on the reasons due to which our results differ from those of earlier work. As explained in some detail at the end of \S~\ref{sec:late_rad_heat}, adopting a strong energy dependence of the energy loss rate $\omega_I=0.5$ (and $x=2$ appropriate only in the highly non relativistic limit), \citet{2018KB_dEdX} find that the late time bolometric light curve decreases slower than obtained here, and does depend on the radioactive energy release rate (as given by Eq.~(\ref{eq:asym_X2}) for $\dot{\varepsilon}\propto t^{-\beta}$). Adopting these results, \citet{2018Kasliwal_IR} use the fact that the late-time IR flux drops faster than would be predicted in this case for $\dot{\varepsilon}\propto t^{-\beta}$ with $\beta\simeq1.3$, to constrain $\dot{\varepsilon}$ and hence the composition. As we have shown here, for weak energy dependence of the energy loss rate ($\omega_I\ll1$), the late time behavior of the bolometric light curve is independent of $\dot{\varepsilon}$ and consistent with the IR observations. Moreover, as noted in the preceding paragraph, estimating the bolometric luminosity based on the IR band luminosity is highly uncertain, and cannot be used therefore to robustly constrain models.

\acknowledgements
E.W. acknowledges support by grants form the Israel Science Foundation, Minerva, and the Israeli Ministry of Technology and Science.
E.O.O. is grateful for the support by
grants from the Israel Science Foundation, Minerva, Israeli Ministry of Technology and Science, the US-Israel Binational Science Foundation, Weizmann-UK,
and the I-CORE Program of the Planning and Budgeting Committee and the Israel Science Foundation.

\appendix

The solution of Eq.~(\ref{eq:Eeq_DL}) for $\epsilon(\tilde{t}_0,\tilde{t})$, with initial conditions $\epsilon(\tilde{t}_0,\tilde{t}_0)=\tilde{t}_0^{-\omega_E}$, is straightforwardly obtained by writing a differential equation for the temporal evolution of $(t^x\epsilon)$. The solution is
\begin{equation}\label{eq:Esol}
  \frac{1}{1+\omega_I}\tilde{t}_0^{(1+\omega_I)(x-\omega_E)}
  +\frac{1}{(1+\omega_I)x-2}\tilde{t}_0^{(1+\omega_I)x-2}
  =\frac{1}{1+\omega_I}\left(\tilde{t}^x\epsilon\right)^{1+\omega_I}
  +\frac{1}{(1+\omega_I)x-2}\tilde{t}^{(1+\omega_I)x-2}.
\end{equation}
For this solution,
\begin{equation}\label{eq:dtdE}
  \left[(x-\omega_E)\tilde{t}_0^{(1+\omega_I)(x-\omega_E)-1}
  +\tilde{t}_0^{(1+\omega_I)x-3}\right]\frac{\partial\tilde{t}_0}{\partial\epsilon}
  =\tilde{t}^{(1+\omega_I)x}\epsilon^{\omega_I}.
\end{equation}

Let us consider now the asymptotic form of $\tilde{t}_0(\epsilon,\tilde{t})$ at $\tilde{t}\gg1$. For $\epsilon\gg t^{-\min[x,2/(1+\omega_I)]}$ we have
\begin{equation}\label{eq:t0h}
  \tilde{t}_0^{x-\omega_E}=\tilde{t}^x\epsilon,
\end{equation}
with $\partial\tilde{t}_0/\partial\epsilon\propto\epsilon^{-1+1/(x-\omega_E)}$. For $\epsilon\ll t^{-\min[x,2/(1+\omega_I)]}$ we have $\tilde{t}_0=Const.\approx 1$ for $x<2/(1+\omega_I)$, and
\begin{equation}\label{eq:t0l}
  \tilde{t}_0^{(x-\omega_E)(1+\omega_I)}=\frac{(1+\omega_I)}{(1+\omega_I)x-2}\tilde{t}^{(1+\omega_I)x-2}
\end{equation}
for $x>2/(1+\omega_I)$. Using Eq.~(\ref{eq:dtdE}), we find  $\partial\tilde{t}_0/\partial\epsilon\propto\epsilon^{\omega_I}$ in the limit $\epsilon\ll t^{-\min[x,2/(1+\omega_I)]}$.

\bibliographystyle{hapj}

\end{document}